The Mastery Rubric for Statistics and Data Science: promoting coherence and consistency in data science education and training

Rochelle E. Tractenberg[1] Donna LaLonde[2], Suzanne Thornton[3]

Reprint Requests:
Rochelle E. Tractenberg
Building D, Suite 207
Georgetown University Medical Center
4000 Reservoir Rd. NW
Washington, DC 20057

rochelle.tractenberg@gmail.com

1. Collaborative for Research on Outcomes and –Metrics; and Departments of Neurology; Biostatistics, Biomathematics and Bioinformatics; and Rehabilitation Medicine, Georgetown University, Washington, DC.
2. American Statistical Association, Alexandria, VA
3. Swarthmore College, Philadelphia, PA

Acknowledgement: There are no actual or potential conflicts of interest.

Running Head: Mastery Rubric for Statistics and Data Science

Abstract

There have been consensus-based publications of both competencies and undergraduate curriculum guidance documents targeting data science instruction for higher education published in the past two years. Recommendations for curriculum features published by the National Academy of Sciences (2018) and the American Statistical Association (2016) may not result in consistency in how across programs will actually be trained. A Mastery Rubric was developed that prioritizes the promotion and documentation of formal growth as well as the development of independence needed for the requisite knowledge, skills, and abilities for professional practice in statistics and data science (SDS). The Mastery Rubric-driven curriculum can emphasize computation, statistics, or a third discipline in which the other would be deployed; or, all three can be featured: the MR-SDS supports each of these program structures while promoting consistency with international, consensus-based, curricular recommendations for statistics and data science, and allows "statistics", "data science", and "statistics and data science" curricula to consistently educate students with a focus on increasing the learners' independence. The Mastery Rubric construct integrates findings from the learning sciences, cognitive and educational psychology, to support teachers and students through the learning enterprise. The MR-SDS will support higher education as well as the interests of business, government, and academic work force development, bringing a consistent framework to address challenges that exist for a domain that is claimed to be both an independent discipline and part of other disciplines (including computer science, engineering, and statistics). The MR-SDS can be used for development or revision of an evaluable curriculum that will reliably support the preparation of early (e.g., undergraduate degree programs), middle (e.g., upskilling and training programs), and late (e.g., doctoral level training) practitioners.

Introduction

There is currently a wide range of definitions of the construct, or discipline, of "data science". What does *not* vary is that the definition includes statistics (and/or mathematics), computing, and some area in which these two must be integrated. This manuscript is grounded in historical efforts to broaden the field of "statistics" so that it can comfortably accommodate both experimental- and discovery- based evidence, particularly featuring the scientific method - either for science or for logical and evidence-based determinations. We assume that data science, as a discipline, must involve data (at any scale) *and science*, defined as "knowledge or a system of knowledge covering general truths or the operation of general laws especially as obtained and tested through scientific method." (Merriam Webster; see also Wu, 1997; Cleveland, 2001; Leek 2013; Donoho 2017). Two core assumptions of this project are:
1. Data science is an extension of applied statistics (Wu, 1997; Cleveland, 2001; Leek 2013; Donoho 2017); and
2. In order to merit the word "science", data science is statistics that is computationally intense and applied to solving problems in a scientific way, i.e., reproducibly and rigorously.

These assumptions behind the definition of "data science" are adapted from those integrated into the definition of "bioinformatics" as: "…the use of computer and information technology, along with mathematical and statistical analysis for gathering, storing, analyzing, interpreting, and integrating data to solve problems" for a given (scientific, industry, business, or government) domain (after Magana et al. 2014; p. 607). Some have labeled "bioinformatics" as "data science for biology" (e.g., Bettinger & Bay, 2023); by featuring an "area of specialization", we seek to ensure that data scientists who practice in government and business settings are fully included.

Clearly, the definition, and assumptions, make many, if not most, statisticians into data scientists (Wu, 1997; Cleveland, 2001; Leek 2013; Donoho 2017). However, these assumptions may also challenge undergraduate programs that currently use the label "data science" to demonstrate (or integrate) the development of scientific reasoning and skills in their students. In order to facilitate coherence and consistency in education and training in the domain of "data science", and to clarify associations between the constituent disciplines, we leverage a curiculum development and evaluation tool called the Mastery Rubric. The Mastery Rubric construct was created in 2005, a rubric that is specifically developed for a curriculum rather than for one assignment. A core feature of a Mastery Rubric for an undergraduate curriculum is the consideration of how an undergraduate and/or graduate program might prepare practitioners in the domain (e.g., Tractenberg et al. 2016; Tractenberg et al. 2019).

A Mastery Rubric makes the developmental trajectory of the instruction that a curriculum is intended to deliver completely explicit (Tractenberg, 2017). Like a standard rubric, every Mastery Rubric is based on curriculum goals, namely the knowledge, skills, and abilities (KSAs) that are the purpose of the educational program. The KSAs make up the rows of the Mastery Rubric. Like a typical rubric, the Mastery Rubric also contains columns that enable the description of each of those learning goals, as they would be performed by individuals who function at different levels. A traditional task-specific rubric describes performance from worse to best, characterizing the work that would merit the grade that the column represents. In a Mastery Rubric, the columns describe recognizable stages of development - so that by observing how any individual performs each KSA, it can be determined whether someone functions at an earlier (or less expert/less independent), at a midpoint, or at the "top" or most sophisticated end of the developmental trajectory. A Mastery Rubric describes *how* the person at a given stage of development performs each particular KSA. The stages are specified up front to capture a meaningful span - and trajectory - for the KSAs throughout a curriculum or developmental path.

No *effective* curriculum can be developed or evaluated without explicit characterization of what the curriculum is intended to deliver (El Sawi, 1996; UNESCO, n.d.*;* Diamond 2008; see also Raffaghelli et al. 2020). Thus, every Mastery Rubric has three dimensions:
1. Each of the elements-knowledge, skills, and abilities (**KSA**s)- that underpin functioning in the domain the Rubric covers;
2. Levels or stages along a developmental trajectory at which an individual would like/be expected to perform; and
3. Performance level descriptors (Egan, Schneider, & Ferrara, 2012; Kingston & Tiemann, 2012) for each KSA, describing how it would be done by someone functioning at each of those levels, or stages.

The **stages** that describe a learner as they change from less- to more- independent, or less- to more- skilled, describe a developmental trajectory to ensure that a curriculum is guiding all learners towards the curriculum's goals. All MRs use the European guild structure (Ogilvie 2014) to describe the developmental stages, with beginners (novices), apprentices, journeymen (who are at the first level of independence) and, for some Mastery Rubrics, Masters. Masters are those who are qualified to take on an apprentice, and although not all of the Mastery Rubrics include a Master Level, this is how the Mastery Rubric got its name (Tractenberg, 2017; Tractenberg, in preparation). The focus on cognitive sophistication is consistent with the emphasis on problem solving, creativity, and innovation that earlier authors intended (e.g., Wu, 1997; Cleveland, 2001; Leek 2013; Donoho 2017). It also allows us to retain the "science" in "data science" by shifting from "problem solving in scientific domains" to "problem solving that is reproducible and rigorous, following the scientific method".

Finally, the intersection of each KSA with the stages requires performance level (i.e., stage) descriptors of that KSA. Reading across any row - i.e., for each KSA - in a Mastery Rubric, the **performance level descriptors** increase in sophistication, complexity, and independence. Reading down any column -i.e., for each stage - in a Mastery Rubric, the performance level descriptors are consistent, even though they describe the performance of different KSAs. The performance level descriptors are the result of applying Bloom's taxonomy to a given KSA, keeping in mind the role of each KSA in the overall function of the domain for which the Mastery Rubric was designed. A core attribute of a Mastery Rubric is that performance level descriptors are concrete and observable, and represent recognizable development of the independence and cognitive sophistication in each KSA.

The creation of a single Mastery Rubric that can accommodate curricular variation in definitions of "data science" (check course listings and/or program descriptions at any two colleges in this listing: https://www.discoverdatascience.org/programs/bachelors-in-data-science/) could facilitate coherence and consistency in the education that is offered across the US. Moreover, it can help to facilitate agreement across instructors about what constitutes "competence" on any KSA (e.g., Fernandez et al. 2012), while also ensuring that the independence in practice attained by the undergraduate in the major is directly related to the further independence to be developed through graduate and post-graduate level training.

Methods

Figure A below outlines how the KSAs for the MR-SDS were created. Development followed a two-phase process, based on the core assumptions and definitions of the domain (SDS). In this case, the core assumption is that the purpose of instruction in SDS is to train scientists (train users of data to do so in a rigorous, reproducible, manner). This leads directly to leveraging the scientific method to identify knowledge, skills, and abilities associated with scientific thinking. Bloom's taxonomy (Bloom et al. 1956) is then leveraged, to ensure that knowledge, skills, and abilities are grounded in observable cognitive behaviors by all learners. Since there exist some competencies lists (e.g., from the US federal job description for "data scientst", or from the ASA 2016 or NAS 2018 curriculum documents), these can be used to qualitatively identify core KSAs. Thus, the first phase is focused on drafting the KSAs to make up the rows of the Mastery Rubric. These activities represent the first two steps in a cognitive task analysis (CTA, Clark et al. 2008; p.580). The five CTA steps are:

1. Collect preliminary knowledge/information;
2. identify knowledge representations and organizations;
3. elicit knowledge;
4. analyze and verify data; and
5. format results.

The figure below shows that KSA development happens during CTA steps 1-2, while the development of performance level descriptors reflect CTA steps 3-4. The Mastery Rubric itself represents CTA step 5.

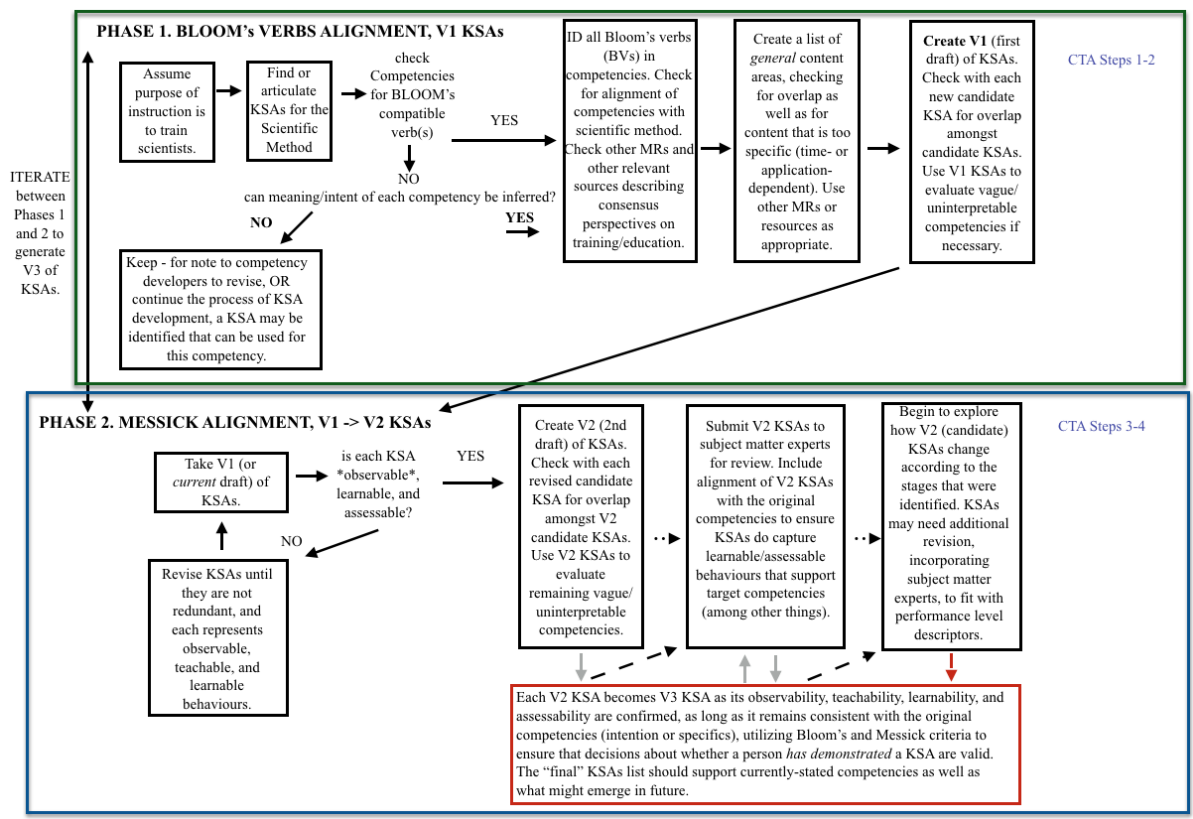

**Figure 1.** Two-phase approach to cognitive task analysis steps 1-4 for KSA extraction. Notes: KSA=knowledge, skills, and abilities; V1= version 1; CTA=cognitive task analysis; ID = identify; BV = Bloom's verbs, i.e., verbs representing observable behaviors reflecting a specific Bloom's

level of cognitive complexity; MR=Mastery Rubric; Messick= following Samuel Messick's three key questions for valid assessment.

**Standard setting**

Standard setting is a collection of formal – reproducible –methods (see Cizek 2012) to generate a reliable description of performance (of a given KSA) at the "conceptual boundary between acceptable and unacceptable levels of achievement" (Kane 1994, p. 433). In every Mastery Rubric, performance level descriptors (Egan, Schneider & Ferrara, 2012; Kingston & Tiemann) are written in a multi-step (iterative) procedure. Generally, the steps are "range-finding", which yields the first draft, relatively general (for the KSA) description of performance of the "minimally competent performer" of each KSA at each developmental stage. The notion of "minimally competent" is essential for the "conceptual boundary between acceptable and unacceptable levels of achievement" - the person who is, in fact, "minimally *competent*" would be on one side of that boundary and the person who is not even minimally competent at the target stage would necessarily be the most-competent performer at the immediately-previous stage.

**Results**

The approach laid out in the cognitive task analysis (Figure 1) was used to generate the KSAs for statistics and data science.

**KSAs**
Based on the fact that "data science" requires both statistics and computing, and the assumption that "data science" requires actual science and there are KSAs for scientific thinking, the 13 SDS KSAs are:
- Prerequisite knowledge -statistics and/or applied math
- Prerequisite knowledge - computational methods
- Prerequisite knowledge – in the discipline where the research questions arise
- Interdisciplinary integration (statistics, computation, and the discipline/context in which these are being deployed)
- Define a problem based on critical review of existing knowledge*
- Hypothesis generation *
- Experimental design *
- Identify or collect data that is relevant to the problem/task/project *
- Identify and use appropriate analytic & computational methods *
- Interpretation of results/output *
- Draw and contextualize conclusions *
- Communication
- Ethical practice

These KSAs are based on Wild & Pfannkuch, 1999 & Bishop 2000 models for doctoral (innovative) life science; and they are essentially the same as the KSAs for the MR for bioinformatics (Tractenberg et al. 2019-a). Importantly, bioinformatics is a similar field to data science, in that there is a key reliance on computing but without directing that computational acumen towards problem solving in the life sciences, the practitioner would not qualify for the title "bioinformatician".

The two "prerequisite knowledge" KSAs (statistics & computing) include "topical" content, <u>facts,</u> and procedures that define their respective domains. The third prerequisite knowledge KSA (in the discipline in which statistics and computing will be deployed) will include content, and may include procedures specific to answering questions; however these are often statistical or computational procedures. "Interdisciplinary Integration", captures communication, teamwork, and also the integration of reasoning, innovation, and documentation for problem solving in a particular field. Eg., applying data science to solve a biological problem engaging basic/bench science, computational methods, and cancer treatment methods, requires integrating both types of prerequisite knowledge *plus* animal, preclinical and clinical models.

The other KSAs reflect the scientific method -indicated with an asterisk (*), communication, and ethical practice. To the extent that one is advanced in all but the scientific method KSAs, one is not a data *scientist*, by definition. Similarly, if one is advanced in all KSAs except the two prerequisite knowledge KSAs for statistics and computation, one would be a *scientist, but not a data scientist.*

Curricular emphasis (i.e., integration throughout the educational experience and along a specified developmental trajectory) on ethical practice is assumed to be essential to any kind of professional preparation: "Upon entry into practice, all professionals assume at least a tacit responsibility for the quality and integrity of their own work and that of colleagues. They also take on a responsibility to the larger public for the standards of practice associated with the profession" (Golde & Walker, 2006: 10).

**Stages:**

Based on the structure of undergraduate programs and prior Mastery Rubric development projects, the MR-SDS was determined to require six stages: beginner, early and late apprentice (two learning stages), and three levels of journeyman. The three journeyman levels (described below) correspond to the levels of independence that describes someone who IS independent, but differs in terms of the extent of independence with which they solve scientific problems. Those who are independent in the practice of statistics and data science, but are not *independent scientists* – or who do not contribute independently to scientific problems (but, e.g., work independently with the job title "data scientist" in business or government), are at the J1, early journeyman stage. There are two levels of independence beyond this, middle (J2) and late (J3) journeymen whose independent work is specifically in the service of solving scientific problems. People could be "even more independent" than late journeyman, but describing those later stages is much more a function of experience and opportunity, and much less a function of instruction and mentorship, which is why later stages are not contemplated in the MR-SDS. Since the MR as a construct is designed to support curriculum development (as well as evaluation and revision), the late journeyman stage is the final stage in the MR-SDS.

**Beginner (e.g., early undergrad)**
- **Bloom's 1** remembering, understanding. Problems that the Beginner can engage with are well-defined, with solutions already worked out/known. Pattern matching.

**Early apprentice (A1) (eg, late undergrad in the major, early masters')**
- **Bloom's 2, 3**, understand and apply **but only apply what you're told to apply**. Problems that the Beginner can engage with are well-defined.

**Late Apprentice (A2) (e.g., master's student, doctoral student/not candidate)**
- **Bloom's 3–4, early 5** <u>Choose and apply</u> techniques in problems that have been defined (either jointly or by others). Analyze and interpret appropriate data. Identify basic limitations and conceptualize a need for next steps/ contextualization of results with extant literature.

**Early journeyman** (**J1**) Independent <u>practitioner</u> (qualified by experience or training/education, or both). May be able to create new code, methods or techniques, but would not be tasked with developing new methods/approaches to problem solving, nor with identifying and solving new problems in their area of specialization.

**Middle Journeyman (J2)** creative innovator to solve problems. Not (yet) independently engaged in the creative, aspects of statistics or data science (e.g., doctoral candidate - or newer workplace group leader).

- **Bloom's 5, early 6** evaluate (review) and synthesize novel knowledge, but not for evaluation of others as they develop their abilities to integrate data scientific methods into the practice of their profession/at work. Independent expertise in the specific area; confident integration of current technology into that area.

**Late Journeyman (J3) (creative, independent innovator)**

- **Bloom's late 6:** synthesize for new knowledge in their area of expertise, metacognitively-aware capacity to consider relevance of "what works" within this area to *other* areas (e.g*., from* economics *for* government) (higher-level synthesis plus evaluation).

**Performance level descriptors**

Following the method published previously (Tractenberg et al. 2019-a; Tractenberg et al. 2019-b), performance level descriptors were iteratively developed. The MR-SDS appears in Table 1.

Table 1. The Mastery Rubric for Statistics and Data Science (MR-SDS). Dark central line delimits developing (A2) vs independent (J1-J3) performance stages.

| Performance Level: | Beginner | Early Apprentice (A1) | Late Apprentice (A2) | Early Journeyman (J1) | Middle Journeyman (J2) | Late Journeyman (J3) |
|---|---|---|---|---|---|---|
| **General description of <u>data scientist</u>** | Uses code, programs and data without synthesis. Generally understands, but does not question, documentation & workflows. Beginning to recognize that *instructions* do not necessarily represent „best practices". Limited engagement with uncertainty associated with outputs across stages of the workflow. Developing understanding of experimental design paradigms in statistics, computation, and engineering | Beginning to learn how to solve given problems (with software, data, and/or simulation). Growing recognition that documentation and workflows are critical for consistency and reproducibility. Engaging consistently with uncertainty associated with "facts"; deepening understanding of experimental design and problem solving paradigms in their area of specialization. | Reads & understands aarguments made with, or from, data and computationally-intensive analyses of that data; reliably identifies methods (software & programming) for given problems. Chooses & executes correct analysis, not necessarily able to identify several methods that could be equally viable, depending on given research objectives. Qualified as a fluent, but not independent, data scientist who uses computation and statistics as necessary. Generates new questions that might not be motivated by (or even recognized as answerable with) new technology. | Independent <u>practitioner</u> (qualified by experience or training/education, or both). Can create new code, but not new methods/approaches to problem solving in science or their area of specialization. New technology/options are seen as potential new avenues, but formulating and pursuing those new ideas requires mentorship and guidance. | Independent data scientist who uses a variety of methodologies as part of routine practice. Creative innovator to solve problems. Identifies data & technology to align appropriate statistical/ analytical methods to accomplish provided objectives; Does not *independently* pose or formulate novel scientific questions or questions for the area of specialization. Experienced reviewer of relevant technical features of available data science methods. Excellent communication. Practitioner who can integrate technology/ techniques into novel questions in their area of expertise. | Independent data scientist who integrates across methodologies, as needed, to achieve desired objectives and identify decisions, and support their resolution/answer (as needed). Can pose novel questions (scientific or in the area of specialization), alone or on a team/in collaboration. Expert reviewer of relevant technical features of relevant statistical, computational, or integrated work. Expert communication. |
| **Evidence of performance at each level** | **Bloom's 1, early 2:** remember, understand. Problems the beginner can engage with are well-defined, with solutions already known. Work does not generally reflect self-assessment. | **Bloom's 2-3**: understand & apply, **but only what you're told to apply**. Problems the early apprentice can engage with are well-defined. Work reflects some self-assessment when directed to do so. | **Bloom's 3-4, early 5:** *choose* & apply techniques to problems that have been defined (either jointly or by others). Analyze & interpret appropriate data. Identify basic limitations & conceptualize a need for next steps/ contextualization of results with extant literature, technology, | **Bloom's 5-6:** Can create new code, but not new methods/approaches to scientific problem solving. Not (yet) engaged in the scientific aspects of statistics or data science. | **Bloom's 5, early 6:** evaluate (review) & synthesize novel knowledge from the area (statistics, computing, data science, or area of specialization) as they develop *their* abilities to integrate new ideas and technologies into their data science practice. Independent expertise in the area of specialization; confident integration of | **Bloom's 6:** prepared for independent work as an expert statistician or data scientist. Expert design & critical evaluation of experimental paradigms & their results. Self-assesses in their work and encourages all those they work with to develop this skill. |

| | | | or prior results. Seeks guidance to improve self-assessment of their work. | | current data science/ technology into that area. Able to begin to critically evaluate experimental paradigms & their results, without knowing/requiring that there be "one right answer". Consistently self-assesses their work. | |
|---|---|---|---|---|---|---|
| **Performance Level:** | **Beginner** | **Early Apprentice (A1)** | **Late Apprentice (A2)** | **Early Journeyman (J1)** | **Middle Journeyman (J2)** | **Late Journeyman (J3)** |
| **KSA 1/13: Ethical practice** | Exhibits respect for community standards/rules for public behavior and personal interaction. Learning how to recognize, and manifest respect for, intellectual property, professional accountability, and intellectual contributions. | Learning to recognize "misconduct" in the scientific sense and professional senses. Learning to avoid, as well as respond to, misconduct; and the importance of neither condoning nor promoting it. | Learning the principles of ethical professional and scientific (ASA, ACM, as appropriate) conduct. Seeks guidance to strengthen their applications of these principles in their own practice. Learning how to respond to unethical practice. | Engages, with mentorship, in how the ethical guidelines from different relevant disciplines pertain in given situations. Discusses ethical practice of statistics and data science, and decision making, with identified mentors. | Consistently practices statistics and data science in an ethical way, and encourages all others to do so as well. Does not promote or tolerate any type of professional or scientific misconduct. Takes appropriate action when aware of unethical practices by others. | |
| KSA 2/13: **Prerequisite knowledge – statistics (includes statistical inference & data** considerations **& experimental or data collection design).** | Basic knowledge of statistical and machine learning methods, but little-to-no awareness of the uncertainty inherent in experimental designs common in science or business applications. Thinking about the/an other area of specialization (or science) is based on uncritical acceptance of information as "factual" or "true". | Advanced knowledge of statistical methods, & basic knowledge of the statistical and inferential aspects of key modeling and machine learning methods. Self contained statistics/programs are run to answer pre-defined questions. Learning to understand the uncertainty inherent in the scientific method, beginning to question assumptions in the data collection, statistical methods, and/or software & their relevance for a given question (which arise from others). | Statistical thinking integrates both experimental )or, data collection) & technological sources for data & knowledge, as appropriate. Understands the uncertainty inherent in the scientific method, questions assumptions in the data & their relevance for a given scientific question (which typically arises from others, or with others). Experimental design & statistical inference are recognized & exploited with guidance, to answer given scientific questions. Can recognize inconsistencies in | Statistically literate, with functional understanding of the inferential and data considerations that are most relevant in their area of specialization. Independence is specific to the area of specialization, and with topics, methods, and questions from prior experience. Within their specific area of application, can troubleshoot experimental methods (e.g., features of data, study design) independently. Developing the ability to critically evaluate the statistical attributes of the relevant literature. | Recognizes the importance of, and is able to critically evaluate, the relevant literature, & understands historical background, of the relevant system(s) and problems in the area of specialization. Sufficient knowledge of the/an other area of specialization (or science generally) to be able to draw functional conclusions from analytical results. Collaborates with experts to inform the next stages in the experimental or statistical/analytical design process (validating results, follow-up analyses, etc.). | Can make predictions to inform next stages of experimental design process. Evaluates relevant experimental methods that can be applied in any problem. Can generalize statistical expertise to other areas of specialization. Independently leverages statistical methods (with or without computationally intensive methods) to solve problems in the area of specialization that are innovative & move statistical practice, and/or the area of specialization, forward. |

| | | | data/experiments that are identified by others, but cannot troubleshoot experimental methods independently. | | | |
|---|---|---|---|---|---|---|
| **Prerequisite knowledge – computational methods (includes computation-specific statistical inference & considerations of system and test designs)** | Basic knowledge of computational methods; little-to-no familiarity with the popular computational methods utilized in the area of specialization. developing awareness of experimental designs or how these can be used or implemented in computational applications. little-to-no awareness of conventional/ popular computational methods. Thinking about tools, computers, software, & programming is concrete, i.e., no extrapolation and/or abstraction of knowledge about computational methods to other systems, programs, or problems. Can run software or execute code they are given (as appropriate) with precise instructions; cannot write a script or debug/ troubleshoot. | Specific computers, software, tools, & programming are recognized as options for scientific work. Learning how to write & test code, run software, or use tools, as appropriate. Is developing awareness of the variety of computational tools, designs, & resources, but is not able to choose or apply the most appropriate of these for any given question; when choices are made, tools are used uncritically. Developing awareness that computational tools require performer-determined input parameters, but uses the default settings. Learning to read, understand, troubleshoot, and make minor modifications to existing code/scripts. Does not synthesize results or outputs. | Learning to test software & programming approaches for different types of problems. Experimental design & statistical inference using computing & algorithms are recognized & applied, with guidance, to answer given scientific questions. Learning "best practices" for programming, if programming is part of the job. Can write basic code in a given language or run appropriate software, using judgment, but not inventing or innovating. Cannot troubleshoot complex computational methods without guidance. Exploring alternatives to default input parameters across computational tools. Can apply knowledge of tools to interpret their results and output. Seeks guidance in synthesis of results or outputs. | Computationally literate, with functional understanding of the computational and technical considerations that are most relevant to computing practices in their area of specialization. Able to understand and apply new/emerging computing methods. Independence in computing is specific to the area of specialization, and with topics, methods, and applications from prior experience. Within their specific area of application, can troubleshoot computational and other technical aspects (e.g., computing/storage requirements, software requirements) independently. Developing the ability to synthesize, and also to critically evaluate synthesis of, results or outputs. | Recognizes the importance of, is able to critically evaluate, & understands historical background of the relevant data, databases, algorithms, tools, data analysis/ statistical methods & computational resources. Can utilise these & justify trade-offs across methodologies (e.g., which statistical test to apply & what computational methods to use). Collaboratively synthesizes and critically questions analysis results & output from tools. Recognises the iterative nature of experiments (e.g., bench, data analysis, back to bench). Can write code/use tools to accomplish these, but collaborates with domain experts for identifying and articulating problems that are innovative & move the field (computing) forward. | Can develop robust, well-documented, optimized, reproducible code &/or use tools to address problems in the area of specialization; can move away from standard procedures & innovate computationally to accommodate new data types, tools, and techniques as needed. Can generalize new coding languages or software/tools/resources for innovative statistical work or to other areas of specialization. Independently leverages computational methods (with or without statistical methods) to solve problems in the area of specialization that are innovative & move computational practice, and/or the area of specialization, forward. |
| **Prerequisite knowledge-** | Does not recognize that/how area of | Becoming familiar with core problems or | Engaging with existing solutions to core | Literate in the area of specialization, with the | Innovative in the area of specialization. Able to | Identifies, and can develop robust, well-documented, |
| area of specialization (includes measurement properties of key outcomes and indicators) | specialization includes specific data features important for statistics and data science applications. | questions, and specific data features relevant in, the area of specialization that are important for statistics and data science applications. Able to recognize core/common problems or questions. | problems or questions, and specific data features relevant in, the area of specialization. Learning to critically evaluate these solutions, and consider how existing solutions to existing problems can inform new solutions to these or new problems in the area of specialization. | burgeoning ability to recognize problems specific to the area that might benefit from statistical and computational techniques to derive new solutions. | critically evaluate the field to articulate newly-identified problems that could benefit from statistical and computational techniques. Can critically evaluate potential solutions and collaboratively trouble shoot with expert (journeyman) computing and statistical experts who may not have extensive experience within the area of expertise. | optimized, reproducible solutions to   problems in the area of specialization. Recognizes limitations in, and can move away from, standard solutions in the area, and innovates to accommodate new or different statistical and computational methods and techniques as needed. Can generalize from other areas of specialization, to yield innovations in their own area. Independently leverages qualitative and quantitative methods to solve problems in the area of specialization that are innovative & move practice in the area of specialization, and/or statistical and computational approaches, forward. |
| Integrate interdisciplin-arity | Does not recognize statistics and data science as requiring integration of experimental, contextual (area of specialization), and computational/modeling approaches. Perceives disciplines as separate; integration only occurs when/as directed. Information, ideas & tools that are interdisciplinary are used without question - particularly, whether or not they are appropriate for a specific question or problem. | Beginning to think about the area of specialization as requiring integration of both experimental & computational/ modeling approaches. Recognizes that interdisciplinarity is needed, but does not know how (or when) to do it, & requires direction. Learning to self-direct the integrating process; learning strengths & weaknesses of applying statistics & computational methods to their area of specialization, but not sufficient to question assumptions | Understands that the area of specialization can integrate both experimental & computational/modeling approaches; seeks guidance about how and when to integrate, strengthening their self-direction. Developing an understanding of the strengths & weaknesses of applying statistics & computational methods to their area of specialization. Beginning to question fundamental assumptions from these & other disciplines for any given question from any area (which arises | When directed, will seek to explicitly integrate information across statistics, computing, and their area of expertise. Integration will typically be based on the prerequisite knowledge they are most familiar with (statistics, computing, or their area of specialization). | Collaboratively integrates across relevant disciplines to address, and solve, innovative problems of their area of specialization. Tests multiple avenues to triangulate solutions, with minimal guidance. Ensures that statistics, computing, and their area of specialization are all integrated effectively and efficiently. Recognizes the roles of interdisciplinary teams in the research process, and the importance of integrating interdisciplinarity early on. Works effectively on interdisciplinary teams with minimal guidance. | Formulates innovative problems in their area of specialization that require interdisciplinary solutions. Integrates methods and results to derive and contextualize solutions to problems in their area of specialization. Consistently tests multiple avenues to triangulate solutions, while exploiting relevant findings from other disciplines. Actively builds interdisciplinary teams, as needed. |
| | | from these & other disciplines. | from others, or in conjunction with others). | | | |
| **Define a problem based on a critical review of existing knowledge‡** | Can recognize a problem that is explicitly articulated or similarly concretely given, but cannot derive one. Unaware of the depth and breadth of "the knowledge base" that is or could be relevant for the formulation of a problem. Does not recognize design features (of data collection or analysis) or other evidence as the basis of/support for problem articulation. Does not recognize or appreciate uncertainty or how this affects the formulation of solvable problems. | Developing awareness of the depth and breadth of „the knowledge base" that is or could be relevant for the formulation of a problem. Cannot differentiate gaps in their own knowledge from gaps in "the knowledge base". Developing the ability to recognize that uncertainty may have arisen in the formulation of solutions to problems. | Beginning to use, with guidance, the appropriate knowledge base to address a given problem. Recognizes the need to consider a wider scope of knowledge for alternative solutions to a problem common across contexts or domains. In guided critical reviews, learning to recognize that design features (of data collection or analysis) & evidence base are important to drawing conclusions. Recognizes the role of uncertainty in research, & that reproducibility & potential bias should be considered for every result. | Can leverage a known problem to explore the literature and documentation, in order to fully describe a problem that is not well understood. With direction, can synthesize known literature from a target domain. Without direction, will not effectively critically review knowledge from diverse domains. | Can explore and critically review the relevant knowledge base; will incorporate information from across disciplines if needed. Can collaboratively articulate a problem based on that review. Reviews include assessment of relevance from (potentially) ancillary domains; bias; reproducibility; and rigor; recognizes when appropriate and inappropriate methodology is used. Recognizes when incomplete review is provided (by themselves or by others). Can discern reproducible from non-reproducible results; can identify major sources of bias throughout the knowledge base. | Independently defines and articulates a theoretical or methodological problems based on a critical review of the relevant knowledge base(s) from their own and other disciplines. Knows the hallmarks of questionable research hypotheses & misalignment of testing/statistics with poorly articulated research questions; consistently finds & identifies sources of bias. Articulates when appropriate & inappropriate methodology is used/reported. Critical review, and problem articulation, integrate diverse contexts when appropriate or adaptable for new applications. |
| **(scientific) Hypothesis generation ‡** | When directed, follows instructions to *test* hypotheses; does not generate them and may not recognize them without explication. Uses the default settings of software and other tools, rather than a hypothesis, to guide any analysis. Does not question methods to be used, or assumptions of methods that are used unless instructed to. | When directed, uses the default settings of software, tools, or the GUI to test hypotheses in pre-planned analyses; does not generate testable hypotheses. Does not recognize that hypotheses may be generated & tested within the intermediate steps of an analysis. Developing the understanding that all methods involve assumptions. | With guidance can leverage tools, software, data & other technologies (GUI/programming) to test hypotheses. Hypothesis generation possible in highly concrete & fully parameterized problems; developing the ability to identify whether a given hypothesis -including one of their own- is testable vs. not. Developing the abilities to identify, & plan to address, assumptions that different | Learning to generate hypotheses based on either the data or the technology, but not their combination/ synthesis. Fully recognizes that a scientific hypothesis and a statistical null hypothesis are fundamentally different entities    . Able to formulate scientific hypotheses for given problems. | Collaboratively integrates hypothesis generation into the consideration of data & analysis options. Can formulate hypotheses for problems that they identify themselves. Seeks appropriate guidance in the synthesis of data and technology to generate novel, testable hypotheses. Considers the process of hypothesis generation and testing to be iterative when this is appropriate. Hypothesis generation is done with consideration of reproducibility and potential for bias, and | Independently generates testable hypotheses that are scientifically innovative as well as feasible (possible for economic reasons, time, impact, etc). In their own & others' work, recognizes that, & articulates how, hypothesis generation from planned & unplanned analyses differ in their evidentiary weight & their need for independent replication. Fully explores all relevant knowledge base(s) to support rigor and reproducibility, and to avoid bias, in the generation of hypotheses. |
| | | | hypotheses necessitate. | | takes into account the most clearly relevant literature; recognizes that less-obviously relevant literature may also be informative for hypothesis generation. | |
| **Experimental Design ‡** | Can recognize concrete features of experiments that are described/given if they match basic design elements (e.g., dependent, independent variables) but cannot derive them if they are not present. Cannot design data collection or experiments. Does not recognize covariates or their importance in analysis or interpretation. Does not recognize the importance of design, data collection, data quality, storage/access, analysis, and interpretation to promote rigor and reproducibility in experimental design. | Can identify salient features of experiments that are described/given if they match previously encountered design elements but cannot derive them if they are not present. Recognizes covariates if mentioned, but does not require formal consideration (or justification) or evaluation of covariates. Does not recognize that one experiment alone cannot adequately address meaningful questions in their area of specialization. Understands that experimental design involves the identification, gathering, storing, analyzing, interpreting, & integrating of data and results. | Can match the correct data collection design to the instruments & outcomes of interest. May include/exclude covariates, or other design features, "because that is what is done" without being able to justify their roles in the hypotheses to be tested. Developing the understanding that weak experimental design yields weak data and weak results. Needs assistance in conceptualizing covariates & their potential roles in the planned analyses. Beginning to recognize that, and can explain why, just one study is usually insufficient to answer a given research questions/solve problems in their area of specialization adequately. Follows templates for the identification, gathering, storing, analyzing, interpreting & integrating of data. Learning to consider reproducibility and rigor in experimental design, and to question templates concerning these concepts. | Can apply a known experimental design to a given problem, and extrapolate this experience to apply known experimental designs to similar but unknown problems. Does not critically evaluate alternative design options, but can critically evaluate the outputs to determine if the solution/results are consistent with prior knowledge or expectations. Will apply sensitivity analyses as directed. May utilize discovery methods rather than designed data collection or generation methods without recognizing the need to confirm results in replicated independent contexts. | Recognizing that explicit attention to experimental design will result in more informative data, designs experiments in consultation with experts in content and statistics. These experiments may include power calculation considerations, if relevant; modeling requirements; measurement/sampling error & missing data. Collaboratively designs experiments that address the need for reproducibility & sensitivity analysis. Learning to conceptualize pilot studies & sensitivity analyses. Learning to adapt problems so that hypotheses can be generated and made testable via experiments. | Can independently design appropriate & reproducible experiments & other data-collection projects, using methodologies that are aligned with the testing of specific hypotheses. Consistently identifies & justifies choices of instruments & outcomes (and covariates if relevant). Collaborates with experts as needed on appropriate use of advanced methods, including accommodating measurement & sampling error, attrition (if needed) and modeling requirements; can adapt complex problems so that hypotheses can be generated and made testable via experiments. Understands & can exploit the strengths & weaknesses of experimental design, data & modeling approaches with respect to the problem under consideration. Uses pilot studies & sensitivity analyses appropriately. |
| **Identify data that are relevant to the problem, task, or project** | Uses data, as directed. Does not find relevant data; cannot describe what makes data or a given data resource "relevant" to a given problem. | Correctly uses data that are provided or can follow a script/"recipe" to obtain (access, manage) relevant data to which they are guided. Cannot determine whether a given data-set or type is relevant for a given problem, but is developing an awareness that not all data are equally relevant, or equally well suited, to all research questions. Developing awareness of the features of data/data resources that constitute "relevance", and that these features, including missingness, must be assessed before choosing data to use. | Can initiate a search for data & will ask if uncertain about the relevance for any given problem. Learning how to identify, & evaluate strengths & weaknesses of, data resources, to determine whether a given data-set or type is relevant for a given problem; and with guidance, how to leverage these to address given research problems. Learning how reproducibility can be affected by the choice (and features) of data. Recognizes the relevance of missingness, will deploy imputation if needed and identified by a mentor. | Can describe, and test, data in order to document fitness for purpose. Will ask for support or input if relevance for problem, task, or project is in question. Able to detect irregularities in data - e.g., diagnose failures to meet assumptions required for methods. Able to fix/accommodate such data when assumptions cannot be met. Recognizes missing data patterns, correctly chooses appropriate handling method (e.g., exclusion, imputation, etc). Able to comprehend ideal data vs realistic data options. | Collaboratively identifies relevant data resources. Understands the relative strengths & weaknesses of data sets and types for addressing their specific problem. Learning to address and formulate testable questions or problems (based on recognized gaps in the knowledge base) utilizing relevant data resources. In their own & others' work, recognizes that, and articulates how, choices for data (collection or use) require assumptions & justification, and must yield reproducible result | Identifies or designs data that are directly relevant to a problem of their own or others' devising. Consistently identifies & evaluates strengths & weaknesses of a variety of data resources that can address a problem or help to formulate it more clearly; recognizes if the necessary data do not yet exist. Justifies the relevance of any given data set to a problem in terms of their individual strengths & weaknesses. Articulates hypotheses, and designs experiments, that leverage strengths in the data; includes triangulating data or results to address weaknesses in the data. Identifies whether or not data appropriate to the specific question were used when reviewing proposals, protocols, manuscripts, and/or other documentation describing data, & research results. |
| **Identify & use appropriate analytical and computational methods** | Uses methods that are provided and in a given order; assumes there are no options for deviation from the given steps/methods. Does not recognize how/when decisions about methods, or steps in solving a problem/answering a question, are appropriate or inappropriate. Does not identify relevant methods; cannot describe what makes a method "relevant" to a given problem. Unaware that methods and software have default settings. Does not question propriety, assumptions, or order of methods that are employed; focus is on the superficial attributes of given methods and protocols. Relies on explicit directions, developing familiarity with formal help documentation. | Uses given methods, as directed, and learning about the decision points and cues for where alternative methods or steps may be important. Beginning to attend to decision points, learning how to make these decisions and run basic checks about appropriate alternatives and their impacts on the resultoutcome. Uses formal, learning to use informal, help documentation. Learning to recognize pros & cons of methods/software, but cannot yet effectively compare, evaluate, or rank them. Becoming aware of the default settings of software or methods and their effects on results; and beginning to explore & inquire about tailored settings. Understands that more than one method/software may be available to deal with a given problem or data type, but can't choose effectively. Learning about similarities and differences across methods, and that choices (particularly of multiple methodologies for one question) should leverage independence of methods to support reproducible results. | Can identify methods, software, and problem-solving paradigms that are relevant for a given problem; seeks guidance about the best approach. Uses formal and informal documentation fluently. Learning to evaluate/rank and justify alternative methods in terms of general features of their efficiency & relevance for the given question. Beginning to recognize problems or situations that require many choices and decisions. With guidance, seeks to identify and implement appropriate approaches to address given research problems. Learning how reproducibility can be affected by the choice and implementation of methods; including independent replication of essentially the same method vs. independent replication using diverse methods. | Will identify analytical and computational methods that are appropriate for a given data set. When seeking new approaches, will incrementally explore new methods that differ from known methods by minimal features (e.g., changing non-parametric for parametric methods). Will seek input from others if known methods do not have straightforward alternatives (e.g., binning continuous variables to generate categorical ones). | Collaboratively considers the knowledge base, and features of the relevant data and analysis options, in identifying the most appropriate approach(es) to tackle a question. Uses appropriate analytic methods and tests whether choices affect mid-point or end-point decisions or inferences, recognizing, and taking advantage of the fact that these may represent distinct approaches to the same problem. Knows when & how to control False Discovery Rates (FDR) to promote reproducible results across methods. In their own & others' work, recognizes that, and articulates how, choices for methods, pipelines, and workflows require assumptions & justification, and must yield reproducible results. | Recognizes if/when the necessary methods to tackle a question or problem do *not yet* exist. Can adapt at any point in a series of steps to ensure that decisions about methods or techniques will affect decisions and inferences are not affected by those decisions. Documents deviations from protocols, and can construct well-reasoned arguments and documentation to ensure appropriate interpretation of the end results, particularly when alternative methods or decisions result in conflicting outcomes. Consistently controls FDR to promote reproducible results. Identifies whether or not appropriate analytical methods were used when reviewing proposals, protocols, manuscripts, and/or other documentation describing methods, techniques, and their sequencing, & the impace of these on results and decisions. Can generate or contribute to new computing software; writes formal help documentation. |
| **Interpretation of results/ output** | Confidently (not usually correctly) treats the output of a program as the final/complete result – with no interpretation required - & is unable to independently execute validation & cross-validation; does not understand their importance for correct interpretation of results/output. Uses the p-value to indicate "truth" in statistical analysis. Over-interpretation is typical. Unaware of the importance of FDR controls. Does not seek coherence in/recognize incoherence of their results with the analysis plan; is unable to align methods, results, & interpretation. | Interpretation of results depends on *p*-values, but understanding of *p*-values is incomplete. Learning to recognize that interpretation of output critically depends on methods used and the sequence of steps from which the results are obtained. Developing awareness of FDR controls. Learning that the interpretation of their immediate results could be an interim step in an overall problem-solving context. | Seeks guidance to interpret results/output, including considerations of alignment of methods and results. Understands that the *p*-value represents evidence about the null hypothesis, not the study hypothesis, but does not consistently avoid prioritizing a favored theory over objectivity. Recognizes that, but does not always act as if, very small *p*-values are *not* "highly significant results". Can apply FDR controls, but does so only when reminded/required. Recognizes when the interpretation of their immediate results is an interim step in an overall problem-solving context. | Can independently interpret results/outputs given the context of the problem. May recognize that limitations, e.g., fitness for purpose of a given data set, necessarily mean no firm conclusions can be drawn. | Can discern, based on immediate results together with methods and hypotheses, whether more experiments &/or data processing are required for robust result interpretation; collaboratively uses the appropriate knowledge base & data resources to interpret results; resists 'overfitting' and resists efforts to ensure a pre-specified outcome; & is committed to good-faith efforts to falsify hypotheses. Consistently & appropriately uses FDR controls. | Interprets their and others' results critically and with respect to the analysis plan; seeks/promotes alignment of methods, results, & interpretation. Prioritizes objectivity, and interpretable & reproducible results above any other outcome (e.g., publication or completion of tasks/project), & insists on FDR controls and other sensitivity analyses in all work. Avoids problems that can arise in the interpretation of results, including bias, overfitting, and the efforts to promote a favored hypotheses. Is able to evaluate the quality & appropriateness of procedures, statistical analyses, & models when reviewing papers & projects/ proposals, based on the writers' –and their own - interpretation of results. |
| **Draw & contextualize conclusions** | Does not draw appropriate conclusions from given results; without direction, will not even contextualize conclusions with the protocol that was followed. Not aware of the difference between conclusions about the null hypothesis and those about the research hypothesis. Conclusions may over- or under-state results & be driven by $p$-values or other superficial cues. Does not recognize the importance of identifying and acknowledging methodological limitations, or their implications, for conclusions. Does not or cannot apply rules of logic to scientific arguments and frequently commits logical fallacies when drawing conclusions. | Learning fundamentals of how appropriate conclusions are drawn from results, but may not be able to draw those conclusions from given results themselves. Learning to differentiate between conclusions about the null hypothesis and those about the research hypothesis. Learning why $p$-value-driven conclusions, and the lack of FDR controls, are not conducive to reproducible work. Conclusions are generally aligned with given results, but when multiple methods are used, does not recognize the dependencies among methods that appear to reinforce, but actually replicate, results. Conclusions are neither fully contextualized with the rest of a document (write-up, paper, etc) or study/ experiment/ paradigm (contextualization for *coherence*), nor with the literature (*critical contextualization*). | With guidance, can draw conclusions in their own work that are coherent with the research hypothesis/hypotheses and across the entire manuscript/writeup or project (as appropriate). Learning to critically contextualize results; is able to draw the most obvious conclusions, but struggles to see patterns, or draw more subtle conclusions. Learning that „full" contextualization of conclusions requires consideration of limitations deriving from methods & their applications, and their effects on results and conclusions. Learning to recognize how independence of multiple methods applied to similar data/problems support reproducible conclusions. | Consistently draws conclusions in their own work that are coherent with the research hypothesis/hypotheses and across the entire manuscript/writeup/project (as appropriate). Will critically contextualize results, but context is typically limited to specific attributes of the problem or project. Understands how rigor and reproducibility depend on good-faith efforts to falsify hypotheses, but may not feel confident resisting pressure to draw firm conclusions when results support less firm conclusions. | Can extract scientific meaning from data analysis and knows the difference between statistical significance and significance within their area of specialization. In their own & others' work, seeks competing, plausible alternative conclusions. Can judge the scientific importance of their results, and draws conclusions accordingly. Can draw conclusions and contextualize results with respect to an entire manuscript/writeup in a given project or study, or with literature (as appropriate). Can detect when conclusions are not aligned with other aspects of the work (e.g., introduction/background, methods &/or results, or other experiments in the project). Careful consideration of limitations deriving from the method & its application in a specific study. Sees patterns, & perceives more subtle conclusions than earlier-stage scientists, and collaborates to fully articulate & motivate them. With collaboration, discusses & draws conclusions (in scholarly work or other documentation), including recognizing limitations. | Expert contextualization of results and conclusions with prior literature, across experiments or studies, & within any given document (e.g., manuscript, writeup, etc.). Strives to fully contextualize conclusions in their own work, and requires this in others' work as well. Draws & contextualizes more subtle conclusions than at earlier stages. Can conceptualize new experiments based on the lack of robust and/or defensible conclusions in others' work. Careful consideration of consistency of conclusions with the other parts of their or others' work. |
| **Communic-ation** | Does not communicate scientific information clearly or consistently; is unaware of community | Learning both to recognize the value of clear communication, and about the role of communication in sharing and publishing research, data, code, | Understands the roles of sharing and publishing research, data, code, data management, tools & resources in scientific communication. Seeks | Competently uses technical language to correctly describe what was done, why, & how. Typically, communication is specific to one or two audience types; with | Consistent and proficient use of technical language to correctly describe what was done, why, & how. Sufficient consideration given to limitations, with explicit contextualization of | Expert communicator & reviewer of communication relating to, or including, statistical & quantitative materials. Consistent sensitivity to audience & appropriate interpretation & |

|  | standards for scientific (or appropriate) communication. Generally relies on lay summaries to support their communication; does not recognize how to use original literature to strengthen scientific communication. Does not differentiate appropriate & inappropriate communication according to area of specialization (e.g., policy; business; science or scholarly writing), nor understand the ethical implications of each. | data management, tools & resources. Developing an awareness of community standards for scientific (or other, audience-appropriate) communication, and that these include documenting code, annotating data, and adding appropriate meta-data. Does not adapt communication to fit the receiver. Learning to differentiate appropriate & inappropriate communication according to area of specialization (e.g., policy; business; science or scholarly writing), but does not yet understand that transparency in all communication represents ethical practice, *even if* the desired results are not achieved. | guidance so that their communicating is coherent, accurate, and consistent with community standards. Learning to document code, annotate data, and add appropriate meta-data –and the importance of these (as appropriate given their research/context) for sharing & integration. Learning the importance of adapting communication to fit the receiver, seeking opportunities to practice this. Learning that transparency in all communication represents ethical practice, even if the desired results are not achieved. | guidance, ensures that what is relayed is understandable and relevant for the target audience. Does not ensure that adapted communications do not inadvertently mislead (or simply not-inform) receivers. | results consistently included in the communication of results and their interpretation. Can adapt communication to fit the receiver without ending up misleading anyone (e.g., by dumbing down to the point of incomprehension). Appropriately annotates all data, code, tools, & resources for documentation, sharing, & integration. Communicates on the appropriate level for a target audience, fully documenting code *and* results. Understands that transparency in all communication represents ethical practice. | contextualization of results. Not only ensures communication is appropriate for a target audience, expertly adapting the communication to fit the receiver. Communication is appropriate to support transparency and reproducibility – and thereby, ethical practice - in every context. |
| --- | --- | --- | --- | --- | --- | --- |
| **Notes:** Solid line between A2 and J1 represents the distinction of independent practice (entry into journeyman stages).<br>‡ These KSAs may be especially critical for data scientists who plan to apply their training in scientific contexts; they would receive more emphasis in doctoral programs where independent research is a targeted objective, and less emphasis in curricula for non-science applications and possibly, for undergraduates. | | | | | | |

The MR-SDS accommodates existing guidance for undergraduate programs in statistics and data science. Because of its basis in the learning sciences (Tractenberg, 2017; in preparation), the MR-SDS supports curriculum and instructional development, as well as evaluation and revision of both, in a manner consistent with formal curriculum design considerations like any Mastery Rubric does, as shown in Table 2.

Table 2. How a Mastery Rubric helps curriculum and instructional development, evaluation, and revision (adapted from Tractenberg, in preparation).

| Five phases of curriculum design (Nichols, 2002): | **Within a Curriculum** | **Instruction/ course** |
|---|---|---|
| Phase 1: **Learning Outcomes** (LO) | PLDs provide ordered, concrete verbs for LO writing; supports planning/development and evaluation. Differences in verbs across stages provide criteria for admission and qualification/completion. LOs can be stated in terms of PLDs and stages; supporting articulation of concrete and observable outcomes. Enables identification of redundancy in the curriculum to promote building up sophistication rather than repetition. PLDs can align faculty (e.g., for courses taken in sequence). | Provides ordered, concrete verbs for LO writing to facilitate focus in a given course or lesson. Supports realistic outcomes given time and prerequisites (LOs can be stated in terms of PLDs and stages). Supports self-directed learners' identification of individualized LOs –leading them to, or following from, the course. PLDs can be used to ensure trajectory for courses to be taken in sequence. |
| Phase 2: **Learning Experiences** (LE) | Much of traditional/existing LEs can be retained, if additional (extra curricular) opportunities are created to fill in any gaps. KSAs and PLDs provide structure for flipping classrooms and trying other innovative teaching techniques/LEs. KSAs, PLDs, and stages help guide curriculum revisions/trying new methods. | Focuses instruction and LEs so that realistic LOs can be accomplished with a variety of content. Enables exploration/creativity in LEs as long as they support the specific (target) KSA(s) and levels that learners want/need. |
| Phase 3: **Content** | KSAs are all scientific method driven, and a variety of content (bioinformatics techniques and reasoning for technological and/or scientific innovation) can be used to teach, provide practice, and assess what learners learn and can do. MR provides rationale for instruction in/with new technology/methods (content) without deviating from | Allows demonstration of relevance of any given content to the learners' goals/career stage. Supports the use of diverse content to enable deeper learning (for greater retention). Supports learners' engaging in their own growth (by enabling their selection of specific learning/content). PLDs allow multiple instructional opportunities at targeted, consistently-graded levels (beginner, median, advanced). |
| | KSAs and movement towards independence across stages.<br>Enables instructors and curriculum developers to engage with/integrate multiple – diverse – content guidelines. | Facilitates the 'cloning' of workshops or courses where KSAs are specifically targeted (or documented) while content can vary. |
| Phase 4:<br>**Assessment**<br>(of learners) | PLDs support targeting of instruction and alignment of LOs with assessment.<br>Capitalize on observability of PLD verbs to design assessments that will be aligned with curricular LOs.<br>Sharing the MR with all learners, and encouraging them to understand the alignment of all courses in the curriculum with the curricular LOs allows them to self-direct (as learners) and utilize assessments to gauge their progress along the KSA trajectories.<br>The PLDs allow faculty to focus assessments and to add opportunities for peer evaluation that further LOs as well as increasing opportunities for formative feedback to learners. | PLDs clarify what learners need to demonstrate, so that "satisfaction" or other survey-based assessments can be replaced with assessments that are aligned with LOs. Sharing the MR with learners, and encouraging them to understand the alignment of the course with the LOs allows them to demonstrate how they a) self direct (as learners) and b) can utilize assessments to gauge their progress along the KSA trajectories. |
| Phase 5:<br>**Evaluation**<br>(of impact of curriculum on learners) | Curriculum can be evaluated for whether LOs were/were not achieved for most learners, focusing on KSAs, and possibly developmental stages, where failures of the curriculum to promote LOs occurred (or occurred most often). LOs can also be evaluated for how realistic they were.<br>Sharing the MR (as intended) can support metacognition throughout the curriculum, to promote self-directed learning beyond the end of the curriculum. Curriculum and instruction evaluation can focus on this. | Promotes targeted follow-up (eg, "if you met the course LOs, did you move on/continue to refine at a level since then?")<br>Strengths and weaknesses in courses can be identified and addressed according to whether LOs, stated in terms of PLDs and stages, were a) reasonable for the time and preparedness of learners; and b) sufficiently supported by the LEs. |
| Notes: KSA= knowledge, skill, or ability; PLD = performance level descriptor; LO=learning outcome; LE = learning experience (teaching technique; MR = Mastery Rubric | | |

Finally, contemplating the KSAs of the MR-SDS facilitates coherence in language as well as instruction for statistics, computing, and data science. Considering the three-dimensional definition of data science as comprising prerequisite knowledge in statistics, computing, and an area of specialization, the potential for the MR-SDS to accommodate and support the development of diverse - but coherent and consistent - curricula in "statistics and data science", "applied statistics", "computational statistics", and other programmatic directions is explored in the figures below. The Figures below describe a wide variety of practitioners discernable by their stage in each of the three dimensions: statistics, computing, and area of specialization. For simplicity, Figures 2-4 reflect just three levels of performance "high" (Journeyman), "medium" (Apprentice), and "low" (Beginner) on the MR-SDS KSAs pertaining to each of the three dimensions of data science (statistics and computing and a third area of specialization).

**Figure 2**. Individuals at the extremes of "prerequisite knowledge" (in statistics, and computing, and/or in an area of specialization) will be at Journeyman (independent) levels on the given dimensions.

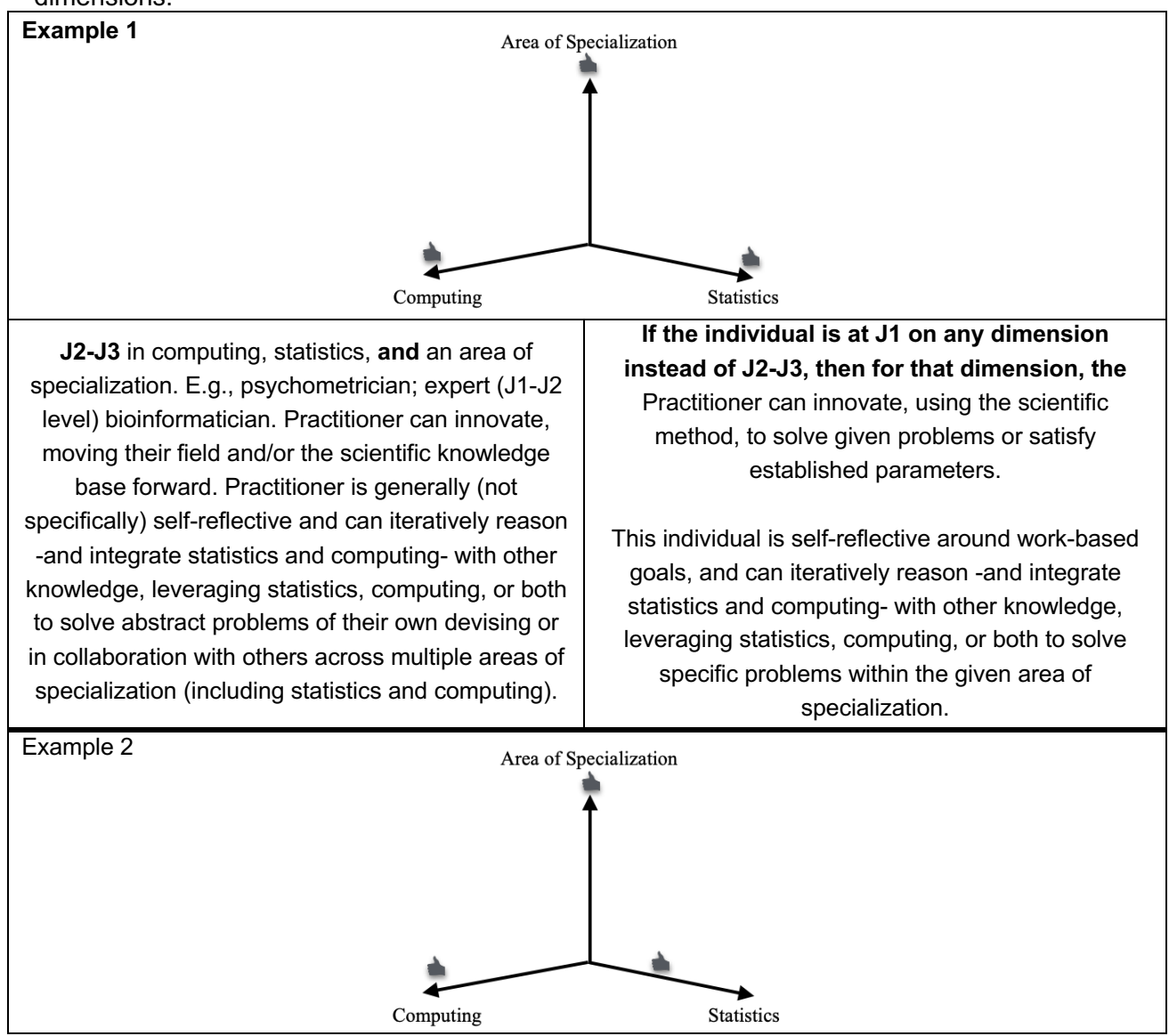

| **Example 1** | |
|---|---|
| **J2-J3** in computing, statistics, **and** an area of specialization. E.g., psychometrician; expert (J1-J2 level) bioinformatician. Practitioner can innovate, moving their field and/or the scientific knowledge base forward. Practitioner is generally (not specifically) self-reflective and can iteratively reason -and integrate statistics and computing- with other knowledge, leveraging statistics, computing, or both to solve abstract problems of their own devising or in collaboration with others across multiple areas of specialization (including statistics and computing). | **If the individual is at J1 on any dimension instead of J2-J3, then for that dimension, the** Practitioner can innovate, using the scientific method, to solve given problems or satisfy established parameters.<br><br>This individual is self-reflective around work-based goals, and can iteratively reason -and integrate statistics and computing- with other knowledge, leveraging statistics, computing, or both to solve specific problems within the given area of specialization. |
| Example 2 | |

| **J2-J3** in computing, and an area of specialization, but **Apprentice** in statistics. E.g., data analytics specialist, bioinformatician with specific focus (e.g., genomics, proteomics). Practitioner can innovate, moving their field and/or the scientific knowledge base forward. Practitioner is generally self-reflective and can iteratively reason -and integrate computing- with other knowledge, leveraging computing to solve abstract problems of their own devising or in collaboration with others across multiple areas of specialization (including computing). Employs a limited set of statistical methods, specific to their area. | **J1** Practitioner can innovate, using the scientific method, to solve given problems or satisfy established parameters.<br>Is self-reflective around work-based goals, and can iteratively reason -and integrate computing- with their knowledge, leveraging computing to solve specific problems within the given area of specialization, possibly using a limited set of statistical methods.<br>**Apprentice** is learning to use a variety of statistical methods with/within the scientific method, how given problems or problem parameters relate to hypotheses and their testing across different statistical methods.<br>Learning to be self-reflective beyond just a specific project/problem, through recognizing the importance of iterative reasoning, and the other knowledge, leveraging statistics, computing, or both to solve specific problems within the given area of specialization. |
|---|---|

Example 3

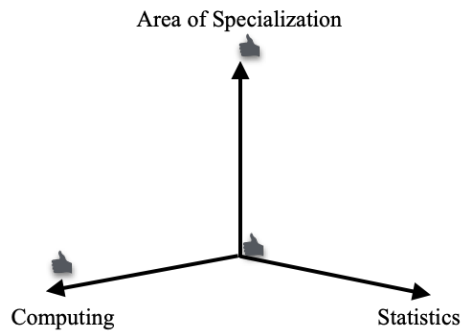

**J2-J3** in computing, and an area of specialization, but **Beginner** in statistics. E.g., data manager; systems architect; data steward. Practitioner can innovate, moving their field and/or the scientific knowledge base forward, but does not leverage statistics effectively (or realize that could be helpful). Is generally self-reflective and can iteratively reason -and integrate computing- with other knowledge, leveraging computing to solve abstract problems of their own devising or in collaboration with others across multiple areas of specialization (including computing). Neither uses, nor recognizes the utility of, statistical methods.

| Example 4 | | |
|---|---|---|
| 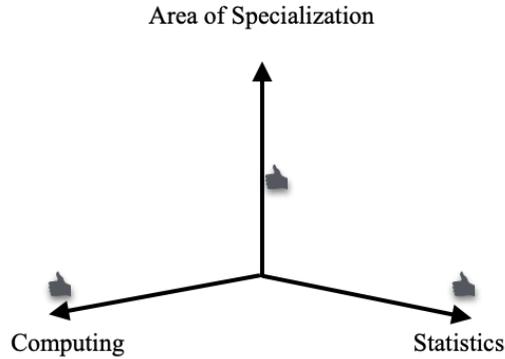 | | |
| **J2-J3** in computing and statistics, but **Apprentice** in an area of specialization, E.g., biostatistician, bioinformatician developing focus (e.g., genetics). Practitioner can innovate, moving statistics, computing, or both forward. Innovates for statistics and computing scientific knowledge base forward. Is generally self-reflective and can iteratively reason and integrate computing with statistical knowledge, leveraging computing to solve abstract problems of their own devising or in collaboration with others across multiple areas of specialization (including computing and statistics). Employs computing and statistics for limited types of problem. | | |
| Example 5 | | |
| 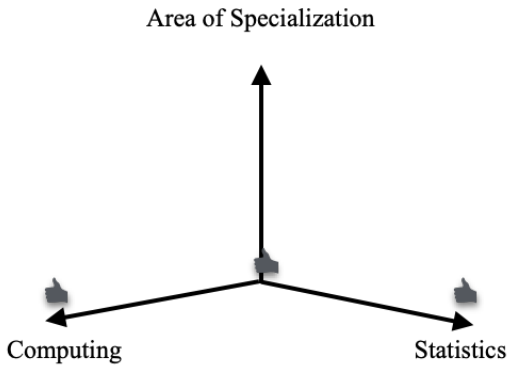 | | |
| **J2-J3** in computing and statistics, but **Beginner** in an area of specialization, E.g., theoretical statistician, computational statistician/statistical computing professional. Practitioner can innovate, moving statistics, computing, or both forward. Is generally self-reflective and can iteratively reason and integrate computing with statistical knowledge, leveraging computing to solve abstract problems of their own devising or in collaboration with others across computing and statistics. Employs computing and statistics for diverse problems without understanding the relevance to any area. | **Beginner** does not utilize the scientific method in an area of specialization. Does not consider/recognize importance of computing and statistical considerations to solve specific problems within any area of specialization. | NB: **Not a data scientist according to ASA, NAS.** (more accurately a "computational statistician") |

Example 6

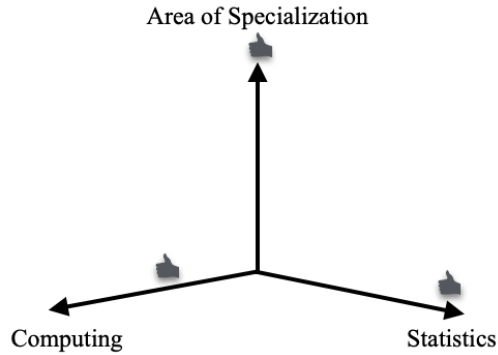

| **J2-J3** in statistics and an area of specialization, but **Apprentice** in computing, E.g., many statisticians and biostatisticians; any PhD scientist relying heavily on experiments, and statistical modeling and methods. Practitioner can innovate, moving their field and/or the scientific knowledge base forward. J2-J3 practitioner is generally self-reflective and can iteratively reason -and integrate statistics- with other domain knowledge, leveraging computing with assistance, to solve abstract problems of their own devising or in collaboration with others across multiple areas of specialization (including statistics). Employs only a limited set of computational methods and workflows. | **J1** Practitioner can innovate, using the scientific method, to solve given problems or satisfy established parameters, utilizing statistics specific to/typically used in the area of specialization. Is self-reflective around work-based goals, and can iteratively reason -and integrate statistics with other knowledge, leveraging known/available computing resources, to solve specific problems within the given area of specialization. **Apprentice** is learning to use computing for the scientific method, how given problems or problem parameters relate to hypotheses and their testing with computers or programming. Learning to be self-reflective through recognizing the importance of iterative reasoning, leveraging **computing**, to solve specific problems using high-level statistics, within the given area of specialization. |
|---|---|

Example 7

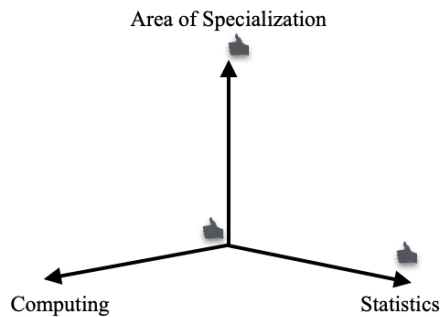

| **J2-J3** in statistics and an area of specialization, but **Beginner** in computing. E.g., any PhD using standard software without questioning the parameters or default settings. Practitioner can innovate, moving their field and/or the scientific knowledge base forward, without taking advantage of any special computational capabilities. | **J1** Practitioner can innovate, using the scientific method, to solve given problems or satisfy established parameters using statistics specialized for their area of specialization. Is self-reflective around work-based goals, and can iteratively reason with other domain knowledge, but does not leverage computing to solve specific problems within the given area of specialization. |
|---|---|

| Is generally self-reflective and can iteratively reason - and integrate statistics - with other knowledge, leveraging statistics to solve abstract problems of their own devising or in collaboration with others across multiple areas of specialization (including statistics -but not computing). | Statistical methods may be limited to what known software offers for the J1 practitioner. **Beginner** does not utilize the scientific method with computing (nor vice versa); does not recognize importance of computing considerations to solve specific problems within the given area of specialization. |
|---|---|
| Example 8 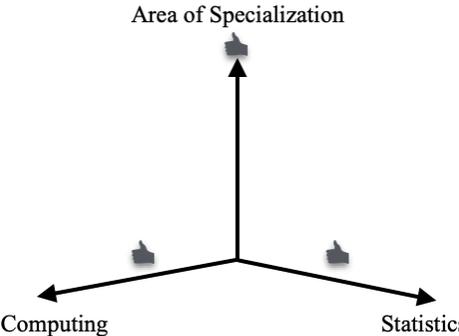 | |
| **J2-J3** in an area of specialization, but **Apprentice** in both statistics and in computing. E.g., typical PhD using (but not developing or contributing to) computationally intense statistical software (somewhat beyond default settings). Practitioner can innovate, moving their field and/or the scientific knowledge base forward, taking advantage of special computational capabilities, but not developing new statistical or computational approaches. Is generally self-reflective and can iteratively reason -and integrate domain-specific statistics and computing- with other knowledge, leveraging statistics, computing, or both to solve abstract problems of their own devising or in collaboration with others across multiple areas of specialization (not including statistics or computing). | **J1** Practitioner can innovate, using the scientific method, to solve given problems in their area, or satisfy established parameters using statistical software. Is self-reflective around work-based goals, and can iteratively reason with other domain knowledge, but does not leverage computing to solve specific problems within the given area of specialization. Statistical methods may be limited to what known software offers. **Apprentice** is learning to use statistics and computing to accomplish key steps in the scientific method, how given problems or problem parameters from the area of specialization can be studied or tested with computers or programming, statistical software (or methods), or both. Learning to be self-reflective about their iterative reasoning, leveraging computing and statistical methods, to solve specific problems within the given area of specialization. |

The examples in Figure 2 can help programs in statistics and data science to clarify how new courses or certificates will enable learners to distinguish themselves and document their performance levels and the domains in which new skills or new levels are aquired. This could be important for initiatives to certify or accredit programs in statistics and data science (e.g., https://www.coursera.org/professional-certificates/ibm-data-science; https://alliancefordatascienceprofessionals.com/).

**Figure 3.** Individuals NOT at the extremes of "prerequisite knowledge".

**Example 1**

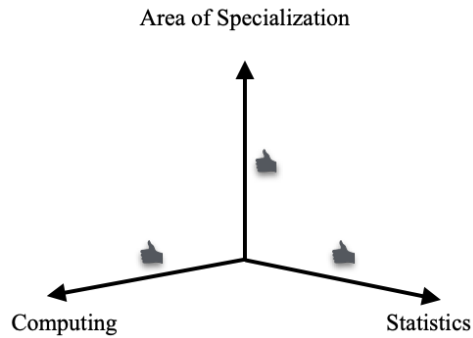

**Apprentice** in computing and statistics and an area of specialization, E.g., junior (or student) biostatistician, bioinformatician without a focus (e.g., genetics), or non-practicing instructor. Practitioner can conceptualize a given solution to a given problem, but does not derive the problem or solution to move a field and/or the scientific knowledge base forward.

Apprentices are not "data science practitioners"; they are not independently self-reflective in statistics, computing, or the area of specialization, only integrates computing with statistical knowledge to solve given, structured, problems. Their jobs or roles may not require this ability to reflect. The Apprentice in statistics and in data science employs known computing and statistics solutions for highly delimited types of problem.

Example **2**

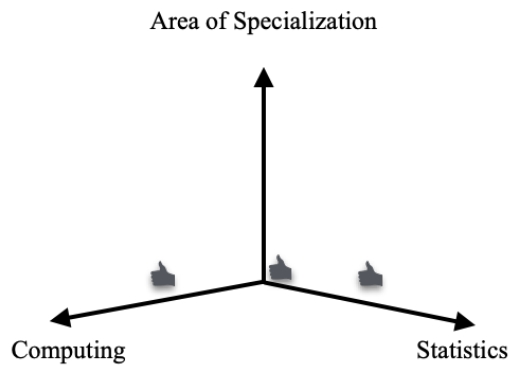

**Apprentice** in computing and statistics, and **Beginner** in an area of specialization, E.g., junior or student biostatistician, bioinformatician without a focus, or non-practicing instructor. Practitioner can conceptualize a given solution to a given problem, but does not derive the problem or solution to move a field and/or the scientific knowledge base forward.

Is not independently self-reflective in statistics or computing, and has no area of specialization, only integrates computing with statistical knowledge to solve given, structured, problems without recognizing applications across domains. Employs known computing and statistics solutions for highly delimited types of problem.

| Example 3 | 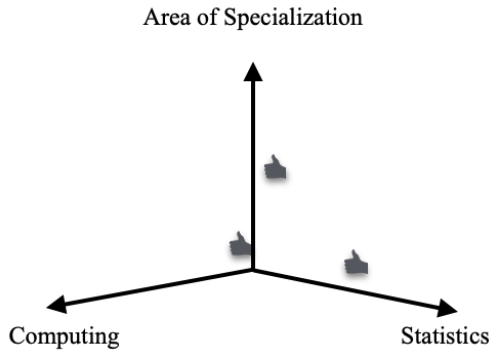 |
|---|---|
| **Apprentice** in statistics and an area of specialization, and **Beginner** in computing E.g., non-practicing instructor in statistics or an area that utilizes statistics explicitly. Practitioner can conceptualize a given solution to a given problem, but does not derive the problem or solution to move a field and/or the scientific knowledge base forward. Is not independently self-reflective in statistics or the area of specialization, only integrates statistical knowledge and the area of specialization to solve given, structured, problems. Employs known statistics solutions for highly delimited types of problem, without considering computational solutions or methods. | |
| Example **4** | 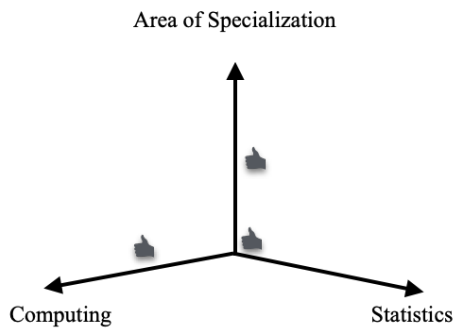 |
| **Apprentice** in computing and an area of specialization, and **Beginner** in statistics, E.g., non-practicing instructor in computing or an area that utilizes computing explicitly. Practitioner can conceptualize a given solution to a given problem, but does not derive the problem or solution to move a field and/or the scientific knowledge base forward. Is not independently self-reflective in computing or the area of specialization, only integrates computing with domain knowledge to solve given, structured, problems. Employs known computing solutions for highly delimited types of problem in the area of specialization, without considering statistical solutions or methods. | |

It is important to note that individuals beyond "Beginner" but not at independent (Journeyman) levels of performance on all three dimensions of "data science" (or applied statistics) as shown in Figure 3: a) can utillize the MR-SDS to find opportunities to develop independence on the articulated KSAs so as to and demonstrate independence; and b) may be expert instructors, leaders, or have other independent level achievement on dimensions outside the MR-SDS scope (e.g., teaching with the Mastery Rubric for the Master Level (Tractenberg 2019) or other certification).

**FIgure 4.** Beginner in statistics and computing and any area of specialization (one example).

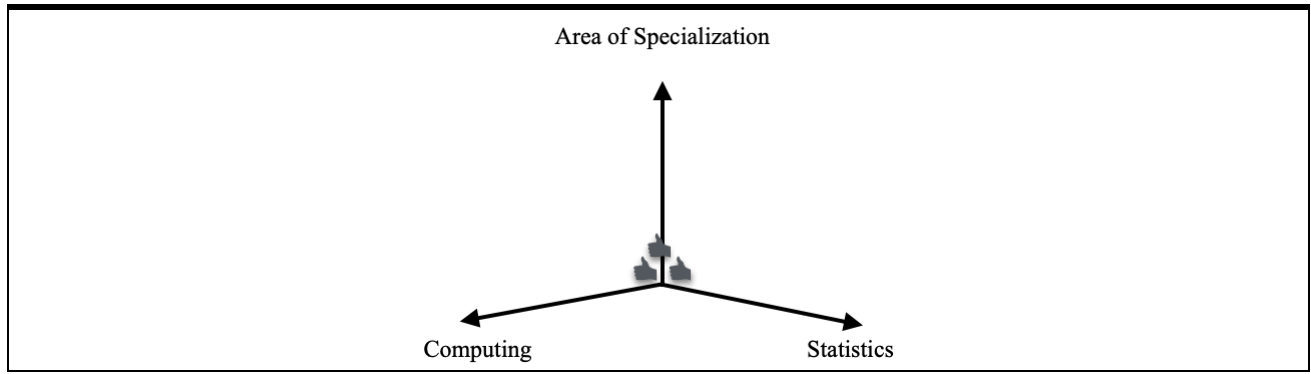

An individual at the Beginner level in all three of the dimensions defining "data science" needs to learn how each of these disciplines formulates questions and problems, as well as the implementation and assessment of solutions. They do not recognize that the scientific method can proceed within each field separately, or in any two, or in all three for a single problem. These individuals do not recognize the contributions to problem solving or scientific hypothesis formation or testing that come from the three domains. They may solve given, structured, problems using the tools that are presented and known, but might not even recognize the utility of tools or methods from one domain for either of the others. Programs that do not generate evidence of functioning by graduates at more advanced levels than this in the three core dimensions of "data science" cannot be considered to be effectively training "data science" practitioners, nor practitioners in applied statistics. Moreover, programs that describe a 'data science competence' that is important to disciplinary or professional practice, e.g., in policy or business, would either specifically designate graduates as "Beginners" if that is the evidence their instruction generates in statistics and computing with respect to their program's area of specialization, or they would seek to incorporate explicit opportunities to move learners beyond the "Beginner" performance levels on all three dimensions. Including one course in "introduction to data science", unless it enables learners to generate this type of evidence would arguably not be providing adequate training or opportunities for learners to demonstrate their acquired performance level(s).

Discussion

The MR-SDS seeks to encourage the development of curricula that promote consistency of learning across undergraduate degrees (in 2- and 4-year programs), but also, true integration across the diverse disciplines that are required for the field to exist. Performance level descriptors are concrete and observable, but flexible. Programs can use the MR-SDS to articulate, with a common vocabulary, what students will learn and how they, their instructors, and employers can recognize that they have learned it. Moreover, learners can utilize the MR-SDS to self-direct their learning, and to document their achievement of target stages on each KSA, using individually-relevant activities and materials.

The MR-SDS has no Master level, because
  a) The MR-SDS is already complex, with 13 KSAs and six stages;
  b) There is a Mastery Rubric describing how anyone can become qualified to take on apprentices (the Mastery Rubric for the Master Level, or MR-ML (Tractenberg, 2019); and
  c) There are many, diverse opportunities for individuals at most stages to provide instruction in any of the KSAs for statistics and data science, so the development of master level qualification does not actually require the achievement of independence for some KSAs.

The MR-SDS accommodates existing guidance for undergraduate programs in statistics and data science, but it provides several other supporting features:

1. Curriculum guidance documents (e.g., National Academies, 2018; ASA 2014) are lists of what should be included in a curriculum but do not address how these competencies or concepts are to be developed – neither over the undergraduate/learner's program nor beyond the end of the undergraduate education. The MR-SDS and particularly, the performance level descriptors that comprise its content, offer guidance on how to promote integration of the prerequisite knowledge in the three defining domains.
2. The development of ethical practice of SDS is described in concrete terms, so that ethical practice can be taught throughout the entire developmental trajectory.
3. Communication, contextualization of results, and integration across disciplines are explicitly included and described, so that graduates of any program will be able to self-direct to ensure they develop those KSAs, *or* the curriculum can be revised to explicitly teach and assess these core KSAs.
4. Programs can choose to emphasize *science* or not – enabling the choice of programs in SDS that actually provide training in the scientific method, or not, as desired.
5. To the extent that computation is emphasized less than statistics in a given curriculum, the program may more accurately be labeled "computational statistics" or just "statistics", while programs that do not focus on statistics would be labeled "computation" or "computer science". The extent of accomplishment and independence on all, or a subset of, these KSAs would be explicit in all programs categorized as "statistics and data science" or "data science".
6. With KSAs that emphasize the role of *science*, long term descriptions of "data science" as essentially applied statistics (e.g., Wu, 1997; Cleveland, 2001; Leek 2013; Donoho 2017) are accommodated. However, programs can feature the scientific method as appropriate for problem articulation and solving within their specific area.
7. The further development of independence as scientists is supported based on the MR-SDS description of earlier achievements on the scientific method KSAs.

The MR-SDS separates interpretation and conclusions as reflected in previous models of the scientific method. Emphasis on these aspects of SDS training will be highly *science* based, so for programs in business or government (non-science) contexts, those KSAs will have less time/effort in the curriculum. However, on the KSAs that are common across programs, students who complete the curriculum will have evidence that they can perform those KSAs at comparable levels. Importantly, while it has been argued that data science requires an explicitly scientific problem solving dimension, we suggest that a subtle shift from "problem solving in scientific domains" to "problem solving that is reproducible and rigorous, following the scientific method" maintains the emphasis on problem solving, creativity, and innovation that earlier authors intended (e.g., Wu, 1997; Cleveland, 2001; Leek 2013; Donoho 2017)

Conclusions

When curriculum and instructional design follow established development guidelines (e.g., Nicholls 2002; Tractenberg et al. 2020), "success" can be characterized, not in terms of completing a series of courses, but in terms of developing the habits of mind and the base of knowledge that can continue to foster excellence in the domain of interest. The intention of using a MR in curriculum development and subsequent instruction and assessment is that claims of learner success and proficiency will be supported with concrete evidence (Mislevy, 2003). When students have moved to the 'independent' side of the rubric, the curriculum can be evaluated in explicit terms – providing concrete characterizations of each student based on work products from the courses, rather than subjective ratings of their satisfaction or student perceptions of instructor effectiveness; nor are characterizations of students as "trained" or "educated" dependent on other, variable (or sample-dependent) methods.

The discipline of data science is new/nascent (National Academies, 2018 p. S-1) but even so, these are not described as wholly *de novo* curriculum development efforts. "Competencies" are included in De Veaux et al., (2017) but learning outcomes are not. Both refer to a "need to include ethics" but neither offers any support in doing so. The KSAs in the MR-SDS can ensure that scientific thinking and aspects of ethical practice (transparency, rigor and reproducibility, as well as how and when to utilize appropriate ethical practice standards, are learned in addition to the constituent topics in math, statistics, computing, and an area in which these are to be applied. Using the MR-SDS would provide the structure –and concrete opportunities for both instructors and students – to integrate course material from the disciplines that make up the statistics and data science curriculum (see De Veaux et al., 2017; Sections 3 & 4). Finally, while any two curricula can be effectively compared on typical criteria such as counts or proportions of students "passing" or graduating; numbers of courses offered; or numbers of passing students, with this definition of success, self-monitoring in this context is limited to "what activity/topic haven't I done yet?" (individual) or "hasn't been offered yet?" (institution). With a MR-based curriculum, monitoring will be more direct in terms of "what can I do/how independently can I do it?" (individual) or "what can my students/graduates do and how independently can they do that?" (institution). The MR-SDS allows internal as well as external evaluation of the functioning of the curriculum (National Academies 2018, Recommendation 5.3). Future work with the MR-SDS will document how courses and curricular recommendations are/can be aligned with it, even if they were not developed with the MR-SDS in mind.

References

American Statistical Association Undergraduate Guidelines Workgroup. (2014). *Curriculum guidelines for undergraduate programs in statistical science*. Alexandria, VA: American Statistical Association. downloaded from https://www.amstat.org/asa/files/pdfs/EDU-DataScienceGuidelines.pdf on 3 December 2016.

Bettinger K & Bay D. (2023, 23 March). *Biology and data science are on a collision course. Here's what you need to know*. Accessed from https://www.weforum.org/agenda/2023/03/biology-data-science-convergence-benefits/ on 4 May 2023.

Bishop G, Talbot M. (2001). Statistical thinking for novice researchers in the biological sciences. In: Batanero C, editor. *Training researchers in the use of statistics.* Granada, Spain: International Association for Statistical Education International Statistical Institute; 2001. pp. 215–226.

Bloom BS, Englehard MD, Furst EJ, Hill WH. (1956). *Taxonomy of educational objectives: The classification of educational goals: Handbook I, cognitive domain. 2nd ed.* New York: David McKay Co Inc.

Cizek GJ. An introduction to contemporary standard setting: concepts, characteristics, and contexts. In: Cizek GJ, editor. *Setting Performance Standards. 2nd ed*. New York: Routledge; 2012. pp. 3–14.

Clark R, Feldon D, van Merriënboer J, Yates K, Early S. (2008). Cognitive Task Analysis. In: Spector JM, Merrill MD, Elen J, Bishop MJ, editors. *Handbook of research on educational communications and technology. 3rd ed*. Mahwah, NJ: Lawrence Earlbaum Associates: pp. 577–593.

Cleveland WS. (2001). Data Science: An Action Plan for Expanding the Technical Areas of the Field of Statistics. *International Statistical Review* 69(1):21-26. https://www.jstor.org/stable/1403527

DeVeaux RD, Agarwal M, Avarett M, Baumer BS, Bray A, Bressoud TC, Bryant L, Cheng LZ, Francis A, Gould R, Kim AY, Kretchmar M, Lu Q, Moskol A, Nolan D, Pelayo R, Raleigh S, Sethi RJ, Sondjaja M, Tiruviluamala N, Uhlig PX, Washington, TM, Wesley CL, White D, Ye P. (2017). Curriculum Guidelines for Undergraduate Programs in Data Science. *The Annual Review of Statistics and Its Application*, **4**, 2.1–2.16. doi: 10.1146/annurev-statistics-060116-053930

Diamond RM. (2008). *Designing and Assessing Courses and Curricula (3E).* San Francisco CA: Jossey Bass.

Donoho D. (September 2017). 50 Years of Data Science, *Journal of Computational and Graphical Statistics*, 26:4, 745-766, DOI: 10.1080/10618600.2017.1384734

Egan K, Schneider C, Ferrara S. Performance Level Descriptors. In: Cizek GJ, editor. *Setting Performance Standards: Foundations, Methods, and Innovations. 2nd ed*. Routledge; 2012. pp. 79–106.

El Sawi G. (1996) Population Education for Non-Formal Education Programs of Out-of-School Rural Youth. http://www.fao.org/3/ah650e/AH650E00.htm

Fernandez N, Dory V, Ste-Marie LG, Chaput M, Charlin B, Boucher A. (2012). Varying conceptions of competence: An analysis of how health sciences educators define competence. *Med Educ.* 46: 357–365. doi:10.1111/j.1365-2923.2011.04183.x

Golde CM & Walker GE. (2006). *Envisioning the future of doctoal education: Preparing stewards of the discipline*. Jossey Bass: San Francisco, CA.

Kane M. (1994). Validating the performance standards associated with passing scores. *Rev Educ Res.* 64: 425–461. doi:10.3102/00346543064003425

Kingston N, Tiemann G. (2012). Setting Performance Standards on Complex Assessments: The Body of Work method. In: Cizek GJ, editor. *Setting Performance Standards: Foundations, Methods, and Innovations. 2nd ed.* New York: Routledge; 2012. pp. 201–223.

Knapper C. (August 2006). *Lifelong learning means effective and sustainable learning: Reasons, ideas, concrete measures.* Seminar presented at 25[th] International course on Vocational Training and Education in Agriculture, Ontario, Canada. Downloaded from http://www.ciea.ch/documents/s06_ref_knapper_e.pdf on 2 Oct 2013.

Knowles MS, Holton III EF, Swanson RA. (2005). *The Adult Learner, 6E*. Elsevier

Messick S. (1994). The interplay of evidence and consequences in the validation of performance assessments. *Educational Researcher* 23(2): 13-23.

Kruger J & Dunning D. (1999). Unskilled and unaware of it: how difficulties in recognizing one's own incompetence lead to inflated self-assessments. *Journal of Personality and Social Psychol*ogy, **77**(6), 1121-1134.

Kuh GD, Ikenberry SO, Jankowski N, Cain TR, Ewell PT, Hutchings P, Kinzie J. (2015). *Using Evidence of Student Learning to Improve Higher Education*; Jossey-Bass: San Francisco, CA.

Leek J. (12 December 2013). "The key word in "Data Science" is not Data, it is Science". Downloaded from https://simplystatistics.org/posts/2013-12-12-the-key-word-in-data-science-is-not-data-it-is-science/ on 5 Dec 2017.

Magana AJ, Taleyarkhan M, Alvarado DR, Kane M, Springer J, Clase K. A survey of scholarly literature describing the field of bioinformatics education and bioinformatics educational research. *CBE Life Sci Educ.* 2014;13: 573–738. doi:10.1187/cbe.13-10-0193

Mislevy RJ. (2003). Substance and structure in assessment arguments. *Law, Probability and Risk* 2: 237-258.

Moseley D, Baumfield V, Elliott J, Gregson M, Higgins S, Miller J, Newton DP. (2005) *Frameworks for Thinking: A handbook for teaching and learning*. Cambridge, UK: Cambridge University Press.

National Academies of Sciences, Engineering, and Medicine. 2018. *Data Science for Undergraduates: Opportunities and Options*. Washington, DC: The National Academies Press. https://doi.org/10.17226/25104

Nicholls G. (2002). *Developing teaching and learning in higher education*. London, UK: Routledge.

Ogilvie S. The Economics of Guilds. *J Econ Perspect*. 2014;28: 169–192. doi:10.1257/jep.28.4.169

Raffaghelli JE, Manca S, Stewart B, Prinsloo P, Sangra A. (2020). Editorial. Supporting the development of critical data literacies in higher education: building blocks for fair data cultures in society. *International Journal of Educational Technology in Higher Education* 17: 50 https://doi.org/ 10.1186/s41239-020-00235-w

Tractenberg, RE. (2017, October 6). *Preprint*. The Mastery Rubric: A tool for curriculum development and evaluation in higher, graduate/post-graduate, and professional education. Published in the *Open Archive of the Social Sciences* (SocArXiv), osf.io/preprints/socarxiv/qd2ae

Tractenberg RE. (2019, August 23). Teaching and learning about teaching and learning: The Mastery Rubric for the Master Level. Published in the *Open Archive of the Social Sciences* (SocArXiv), https://doi.org/10.31235/osf.io/md65v

Tractenberg RE. (Book in preparation). *Practical higher education curriculum and instructional design with the Mastery Rubric.* ( ~350 pages)

Tractenberg RE, Gushta MM, Weinfeld JM. (2016). The Mastery Rubric for Evidence-Based Medicine: Institutional Validation via Multidimensional Scaling. *Teach Learn Med*. 28: 152–165. doi:10.1080/10401334.2016.1146599

Tractenberg RE, Lindvall JM, Attwood TK, Via A. (2020, April 2) Preprint. Guidelines for curriculum and course development in higher education and training. *Open Archive of the Social Sciences (SocArXiv)*, doi 10.31235/osf.io/7qeht.

Tractenberg RE, Wilkinson M, Bull A, Pellathy TP, Riley JB. (2019). A developmental trajectory supporting the evaluation and achievement of competencies: Articulating the Mastery Rubric for the nurse practitioner (MR-NP) program curriculum. *PLoS ONE 14(11): e0224593. https://doi.org/10.1371/journal.pone.0224593*

*UNESCO - International Institute for Educational Planning (n.d.) -Glossary: Curriculum*
https://learningportal.iiep.unesco.org/en/glossary

Wild CJ, Pfannkuch M. Statistical thinking in empirical enquiry. *Int Stat Rev*. 1999;67: 223–248. doi:10.1111/j.1751-5823.1999.tb00442.x

Wu CFJ. (1997). "Statistics = Data Science?" Downloaded from https://www2.isye.gatech.edu/~jeffwu/presentations/datascience.pdf on 16 November 2019.